\documentstyle[12pt,twoside,fleqn,espcrc1]{article}

\newcommand{\AmS}{{\protect\the\textfont2
  A\kern-.1667em\lower.5ex\hbox{M}\kern-.125emS}}

\def\vsig{\mbox{\boldmath$\sigma$}}
\let\vec\bf
\def\SU#1{SU(#1)}
\def\Lam{\Lambda}
\def\Sig{\Sigma}
\def\wave{\simeq}
\def\ie{{\it i.e.}}

\input{epsf}

\title{Roles of Quark Degrees of Freedom in Hypernuclei}

\author{Makoto Oka\address{Department of Physics,
        Tokyo Institute of Technology \\
        Meguro, Tokyo, 152 JAPAN}%
        \thanks{email: oka@th.phys.titech.ac.jp}
        }

\begin{document}
% typeset front matter
\maketitle

\begin{abstract}
The quark model description of the hyperon nucleon forces, especially 
the antisymmetric spin-orbit forces, is studied from the spin-flavor 
\SU6\ and the flavor \SU3\ symmetry point of view.
It is pointed out that the quark exchange interaction predicts strong 
antisymmetric spin-orbit force between the hyperon and nucleon.
\end{abstract}

\section{Introduction}

In this talk, I cover the following four subjects.
\begin{enumerate}
\item Quark model description of the baryon-baryon interaction 
including hyperons{\cite{YITP}}.
\item Antisymmetric spin-orbit force from the \SU3\ flavor symmetry 
point of view{\cite{TO}}
\item Description of weak $\Lam N \to NN$ interaction in the direct 
quark mechanism{\cite{ITO}}.
\item Magnetic moments of light hypernuclei and contributions of 
$\pi$, $K$ exchange currents{\cite{SOS}}.
\end{enumerate}

In this report, I discuss the first two subjects in detail and leave 
the others to references given above.

\section{Baryon-Baryon Interactions}

Recent experimental activities in hypernuclear physics provide us
with high quality data of production, spectra and decays of hypernuclei.
The accumulation of such data accelerates quantitative analyses 
of the strong and weak interactions of hyperons.
Theoretical efforts have been devoted to understanding hypernuclear 
structure and production and decay mechanisms.  There the most 
important ingredient is the hyperon-nucleon interactions.
Several realistic potential models are in use 
widely{\cite{Nijmegen,DF,Julich}}, 
but their 
foundations are not solid and furthermore, there are some 
discrepancies among the models.  For instance,
the strengths of the spin-spin interaction vary significantly 
among the models. It seems urgent to establish the quantitative 
description of the hyperon-nucleon (YN) interactions.

In studying the YN interactions, it is natural to follow the 
description of nuclear force.  The 
long-range part of the nuclear force is explained very well in terms 
of one-pion exchange mechanism, while heavy mesons as well as 
multi-pion exchanges are necessary for the medium range part of the 
nuclear force.  One-boson exchange potential models, in which 
two- (and multi-) pion exchanges are taken into account as the 
$\sigma$ and $\rho$ exchanges, are fairly successful in accounting the 
large amount of data for nucleon-nucleon scattering \cite{Bonn}.
Yet the short-range part of the nuclear force is not fully 
treated, that is, repulsive cores (hard or soft) are introduced 
phenomenologically to 
explain the NN scattering phase shifts around $E\wave 100-200$ MeV in 
the center of mass system.  Indeed, this is the region where the 
internal quark-gluon structure of the nucleon must be considered 
explicitly.

It was pointed out that the quark exchange force between two nucleons 
give significant repulsion at the short distance.  The exchange force 
is induced by the quark antisymmetrization and therefore is nonlocal and 
of short-range determined by the size of the quark content of the 
nucleon.
The most important feature of the quark exchange force is its 
dependence on the spin-flavor symmetry of two-baryon states.
A close analogy is found in the hydrogen molecule, where two 
electrons orbit around two protons.  As the total spin of the 
electrons specifies the symmetry of the spin wave function, the sign 
of the exchange force is determined according to the spin.
The symmetric orbital state is allowed only for $S=0$, while the 
exchange force is strongly repulsive for $S=1$.
Similar state dependencies appear in the quark exchange force, where 
the spin-flavor \SU6\ symmetry determines the properties of the 
exchange interactions.

We applied the quark cluster model description of the short-range 
YN interactions{\cite{OSY}}.  We found that the flavor singlet 
combination of $\Lam\Lam-N\Xi-\Sig\Sig$ has no repulsion 
induced by the quark exchange at the short 
distance.  As this state is known to be favored by the magnetic part of the 
one-gluon exchange interaction (color-magnetic interaction), a bound 
or a resonance state called H dibaryon may exist{\cite{Jaffe,hakone}}.   
On the other hand, most other channels have strong 
repulsion at short distances.
  
Another interesting observation made in ref.\cite{OSY}\ is that the 
$S$ wave $\Sig N$ interaction depends strongly on the total spin and 
isospin.  The $\Sig N$ ($I=1/2$, $S=0$) and $\Sig N$ ($I=3/2$, $S=1$)
states belong mainly to the [51] irreducible representation of the 
spin-flavor \SU6\ symmetry.  This is the representation in which a 
Pauli forbidden state appears in the $L=0$ orbital motion.  The Pauli
principle forbids two baryons to get together and thus gives a strong 
repulsion.  The other spin-isospin states do not belong to this 
symmetry and therefore the short range repulsion is weaker.
This qualitative argument was confirmed in realistic quark cluster 
model calculation of the YN interactions{\cite{OSY,NSF}}.
Recent analyses by Niigata-Kyoto group show that the strong repulsion 
remains after combining the quark exchange interaction with the 
long-range meson exchange attraction{\cite{NSF}}. 
No experimental evidence is yet available to be able to
confirm the strong state dependencies.  
More $\Sigma N$ scattering data are anticipated very much.

\section{Antisymmetric Spin-Orbit 
Forces\protect\footnote{This part of the work has been done in collaboration
with Yoshihiro Tani\protect\cite{TO}.}}

One of the interesting features of the hyperon-nucleon interactions is the 
properties of the spin-orbit force.
The Galilei invariant spin-orbit force consists of symmetric LS (SLS) and 
antisymmetric LS (ALS) terms,
\begin{eqnarray}
   V_{SO} &=& V_{SLS}({\vsig_1}+{\vsig_2})\cdot {\vec L}
  +  V_{ALS}({\vsig_1}-{\vsig_2})\cdot {\vec L} \nonumber\\
   &=&  (V_{SLS}+  V_{ALS})\,{\vsig_1}\cdot {\vec L}
   +(V_{SLS}-  V_{ALS})\,{\vsig_2}\cdot {\vec L}
\end{eqnarray}
Because the ALS operator $(\vsig_1-\vsig_2)\cdot \vec L$
is antisymmetric with respect to the exchange of two baryons, 
$V_{ALS}$ should be zero between like baryons.
In the nuclear force, ALS between proton and neutron breaks the isospin 
symmetry and is classified as 
a type IV charge symmetry breaking (CSB) force.  
Evidence of such a CSB force was
given by measuring the difference between the proton and 
neutron analyzing powers in the $n-p$ scattering experiments\cite{AP-exp}.
The results show that this force is very weak, supporting the isospin 
invariance of the nuclear force.

On the contrary, the ALS forces in the hyperon-nucleon interactions do not 
vanish even in the SU(3) flavor symmetric limit.
For hyperon-nucleon systems, ALS seems as strong as
the symmetric LS part.
If the magnitudes of SLS and ALS are comparable, then  
the single particle LS force for one of the baryons, (ex.\ $\Lambda$) 
inside (hyper)nuclei is much smaller than that for the other baryon (nucleon).
Since recent experiment suggests that the single particle LS force for 
$\Lambda$ might be sizable contrary to the wide belief of vanishing LS
force for $\Lambda$,  it is extremely important to  pin down the
magnitude of the two-body LS force.
Here we study the properties of the YN ALS forces from the SU(3) 
symmetry point of view.

\subsection{\SU3\ Invariance}
\def\rep#1{{\bf {#1}}}

For the octet baryons, the baryon-baryon interactions can be 
classified in terms of the SU(3) irreducible representations given by
\begin{equation}
 \rep8 \times \rep8 = \rep1 + \rep{8_s} + \rep{27} + \rep{10} + \rep{10^{*}}
     + \rep{8_a}
\end{equation}
Among these six irreducible representations, first three,  $\rep1$, 
$\rep{8_s}$, and
$\rep{27}$, are symmetric under the exchange of two baryons and the other three
$\rep{10}$, $\rep{10^{*}}$ and $\rep{8_a}$, are 
antisymmetric{\cite{DF}}.  
Noting that two-baryon states are to be
antisymmetric, and 
that the color wave function for the color-singlet baryons is always symmetric, 
we find that the symmetric (antisymmetric) flavor representations are combined 
only to antisymmetric (symmetric) spin-orbital states.

In baryon-baryon scattering, the ALS force induces the transition 
(mixing) between the spin singlet states ($^1P_1$, $^1D_2$, $^1F_3$, \ldots)
and the spin triplet states with the same $L$ and $J$ 
($^3P_1$, $^3D_2$, $^3F_3$, \ldots).
The flavor symmetries of $^1P_1$ state must be antisymmetric, 
$\rep{10}$, $\rep{10^{*}}$ and $\rep{8_a}$, while
that of $^3P_1$ state is symmetric,  $\rep1$, $\rep{8_s}$ or $\rep{27}$.

If one assumes the \SU3\ invariance of the strong interaction,
different irreducible representations are not mixed. 
Therefore the only possible combination of symmetric and antisymmetric 
representations is $\rep{8_s}- \rep{8_a}$.
We conclude that the ALS in the SU(3) limit should only connect 
$\rep{8_s}$ and $\rep{8_a}$. 

The symmetry structure becomes clearer by decomposing 
the $P$--wave $\Lam N- \Sig N$ ($I=1/2$), as a concrete example,
into the \SU3\ irreducible representations.
The flavor symmetric states read
\begin{equation}
  \pmatrix {\Lambda N \cr \Sigma N\cr} (^3P_1)
      =  \pmatrix {\sqrt{9\over 10} & -\sqrt{1\over 10} \cr
                   -\sqrt{1\over 10} &  -\sqrt{9\over 10} \cr }
        \pmatrix{ \rep{27} \cr \rep{8_s} \cr} 
\end{equation}
while the antisymmetric ones are
\begin{equation}
  \pmatrix {\Lambda N \cr \Sigma N\cr}  (^1P_1)
      = \pmatrix {-\sqrt{1\over 2} & -\sqrt{1\over 2} \cr
                   -\sqrt{1\over 2} &  \sqrt{1\over 2} \cr }
        \pmatrix{ \rep{10*} \cr \rep{8_a} \cr} 
\end{equation}
In the \SU3\ limit, the only  surviving matrix element is
$\langle \rep{8}_a \, ^1P_1 | V| \rep{8}_s \, ^3P_1\rangle$.
When we turn to the YN particle basis, we obtain the following 
relations in the \SU3\ limit.
\begin{eqnarray}
  \langle \Lam N \,^1P_1 | V| \Lam N\, ^3P_1\rangle  &=& 
      -\, \langle \Sig N^{(1/2)} \,^1P_1 | V| \Lam N\, ^3P_1\rangle \\
  \langle \Lam N \,^1P_1 | V| \Sig N^{(1/2)}\, ^3P_1\rangle 
  &=& - \, \langle \Sig N^{(1/2)} \,^1P_1 | V| \Sig N^{(1/2)}\, 
  ^3P_1\rangle
  = 3\,\langle \Lam N \,^1P_1 | V| \Lam N\, ^3P_1\rangle
\end{eqnarray}

On the contrary, the  $\Sigma N$ ($I=3/2$) system belongs purely to the
\SU3\ $\rep{27}$ and therefore the ALS matrix element vanishes in
the \SU3\ limit.
\begin{equation}
\langle \Sig N^{(3/2)} \,^1P_1 | V| \Sig N^{(3/2)} \, 
^3P_1\rangle =0 
\end{equation}

These relations come only from the \SU3\ symmetry and is general for 
any ALS interactions regardless their origin.
It is extremely interesting to note that (1) the ALS for $\Sigma 
N^{(1/2)}$ is much stronger than and has different sign from that 
for $\Lam-N$,  (2) the coupling of $\Lam N - \Sig N^{(1/2)}$ is also strong,
and (3) the ALS for $\Sigma N$ depends strongly on the isospin or
the charge states.

\subsection{Quark Cluster Model Potential}

The quark substructure of the baryon must play significant roles at 
short distances in the baryonic interactions{\cite{OY}}.
The quark model symmetry is especially simple when two baryons sit 
on top of each other.
For $L=1$  states, the six-quark orbital state takes [51] symmetry, 
which can couple only to the [42] \SU6\ representation due to the 
Pauli principle.  Namely the [6] symmetric \SU6\ states are not allowed 
to couple to [51] orbital configuration.
The $I=1/2$ $\Lam N -\Sig N$ states at $R\to 0$ are given 
in the SU(6)/SU(3) symmetry basis by
\begin{eqnarray*}
   |S=0\rangle &=& |[42] \rep{8_a}\rangle +  |[42] \rep{10^*}\rangle \\
   |S=1\rangle &=& |[42] \rep{8_s}\rangle +  |[42] \rep{27}\rangle 
\end{eqnarray*}
Again the ALS in the \SU3\ limit can connect only the octet states,
$[42] \rep{8_a}$ and $[42] \rep{8_s}$,  and
the ratio of the ALS interaction in the particle basis,
$\Lam N -\Sig N$ should follow eq.(3).

The quark-quark potential due to one-gluon exchange 
gives a $q-q$ spin-orbit interaction.
Its contribution to the YN ALS forces are calculated in the above 
mentioned limit ($R=0$) {\cite{Morimatsu}}.  
They are compared with the corresponding
potential of the ordinary spin orbit force between the $^3P_1$ states
in Table \ref{tab:QCM}.

The results show that the ALS forces due to the quark exchange force
are as strong as the ordinary LS force of the same origin.
Especially, the $\Sigma N$ ($I=1/2$) feels a stronger ALS 
force between $S=0$ and $S=1$, than the ordinary LS force
between $S=1$ states.
This is a very interesting result.  The quark exchange ALS force might 
play dominant role in the YN forces, as was already suggested and 
demonstrated by Kyoto-Niigata group{\cite{NSF}}.

\begin{table}[hbt]
% -----------------------------------------------------
% adapted from TeX book, p. 241
%\newlength{\digitwidth} \settowidth{\digitwidth}{\rm 0}
%\catcode`?=\active \def?{\kern\digitwidth}
% -----------------------------------------------------
\caption{ALS and LS matrix elements at $R=0$ normalized by the
overlapping matrix element.}
\label{tab:QCM}
%\begin{tabular*}{\textwidth}{@{}l@{\extracolsep{\fill}}|rr|rr}
\begin{tabular}{l|rr|rr}
\hline
&\multicolumn{2}{c|}{$\langle ^1P_1| V_{ALS}|^3P_1\rangle$}&
 \multicolumn{2}{c}{$\langle ^3P_1| V_{LS}|^3P_1\rangle$} \\
&\SU3\ limit&\SU3\ broken&\SU3\ limit&\SU3\ broken\\
\hline
($I=1/2$)&(MeV)&(MeV)&(MeV)&(MeV)\\
$\Lam N \leftarrow \Lam N$& $37$ & $32$ & $-74$ & $-55$ \\
$\Lam N \leftarrow \Sig N$& $88$ & $77$ & $33$ & $29$ \\
$\Sig N \leftarrow \Lam N$& $-37$ & $-29$ & $33$ & $29$ \\
$\Sig N \leftarrow \Sig N$& $-88$ & $-79$ & $22$ & $22$ \\
\hline
($I=3/2$)&&&&\\
$\Sig N \leftarrow \Sig N$& $0$ & $1$ & $-95$ & $-94$\\ 
\hline
\end{tabular}
%\end{center}
\end{table}

Table \ref{tab:QCM}\ also shows the potential values when the 
\SU3\ symmetry is 
broken by the mass difference of the strange quark and the $ud$ quarks.
The effects of the symmetry breaking are not so large that the results
are essentially the same.
Thus the above \SU3\ relations of the ALS matrix elements remain 
valid qualitatively, \ie, 
$\langle \Lam N \,^1P_1 | V| \Sig N^{(1/2)}\, ^3P_1\rangle$ and 
$\langle \Sig N^{(1/2)} \,^1P_1 | V| \Sig N^{(1/2)}\, ^3P_1\rangle$ 
are large, while 
$\langle \Sig N^{(3/2)}\,^1P_1 | V| \Sig N^{(3/2)}\, ^3P_1\rangle$
almost vanishes.

\subsection{Meson Exchange Potential}
For meson exchange interactions exchanged mesons considered 
are either in the flavor singlet or octet representation
(because they are $q\bar q$ states).  
The SU(3) factor for the (meson $M^a$)--(baryon $B_i$)--(baryon $B_j$)
coupling, $T^a_{ij}$, has three choices, $\delta_{ij}$ 
for the flavor singlet meson ($a=0$) and $F_{aij}$ or $D_{aij}$ 
for octet mesons ($a=1-8$), 
where $F$ and $D$ are symmetric and antisymmetric SU(3) 
structure constants, respectively.  
Then the SU(3) invariant potential is proportional to 
\[
 \sum_{a=1}^{8} \,( T^a_{ij} \cdot T^a_{lm} + \hbox{exchange term})
\]
Thus possible antisymmetric coupling is of the form
$ (F_{aij}\cdot D_{alm} - D_{aij}\cdot F_{alm})$.
This term, however, vanishes because the ratio of the 
$f$ and $d$ couplings is fixed for each meson 
without depending on the choice of
baryons ($ijlm$). 
One exception is for the vector and the 
tensor couplings in the vector meson exchange force.
According to the vector meson dominance, $F/D$ ratio for the vector 
and the tensor couplings are in general different and then terms 
like $(g_1 f_2 - f_1 g_2) (\vsig_1-\vsig_2)\cdot \vec L$
will survive, where $g_k$ ($f_k$) is the vector (tensor) coupling 
constant of a vector meson to a baryon $k$ ($k=1$ or 2).

One can confirm the above symmetry consideration by a look on the ALS 
potential term in the Nijmegen potential{\cite{Nijmegen}}, for instance.
There the SU(3) symmetry is broken by differences among the baryon 
masses that cause factors like $(M_Y^2 - M_N^2)/4M_YM_N$.
The equal baryon mass $M_Y=M_N$ in the SU(3) limit kills most terms,
leaving 
\begin{equation}
 V_{ALS} (r) = {g_{Y} g_{N}\over 4\pi}\, {m^3\over 2M^2}\, 
\xi(x)
\left[\left({f\over g}\right)_{N} - \left({f\over g}\right)_{Y} 
\right] \, ({\vsig_1}-{\vsig_2})\cdot {\vec L}
\end{equation}
where $m$ is the meson mass, $x=mr$ and $\xi(x)$ is a radial function defined by
\begin{equation}
  \xi(x)=\left({1\over x} + {1\over x^2}\right) \, {e^{-x}\over x} \quad.
\end{equation}

In the \SU3\ limit, the coefficients of ALS force due to exchanges of
the vector mesons, $\rho$, $\omega_8$ and $K^*$ are given for the \SU3\ 
basis, and also for the particle basis in Table \ref{tab:OBE}.
The common factor 
$ {1\over\sqrt{5}} \, fg (\alpha-\beta)$
contains $\alpha= F_g/(F_g+D_g)$ for the vector($g$)-coupling
and $\beta= F_f/(F_f+D_f)$ for the tensor($f$)-coupling.
All the matrix elements vanish when $\alpha=\beta$.
Thus it is critical for ALS to have different $F/D$ ratios for the 
vector and tensor couplings.

%%%%%%%%%%%%%     VERSION  2        %%%%%%%%%%%%%%%
\begin{table}[hbt]
\caption{Coefficients of the ALS force due to the vector meson exchanges. 
The four numbers in 
each raw are the contributions of $\rho$, $\omega_8$ and $K^*$ 
vector mesons and their sum, respectively, with a common factor, 
$fg(\alpha-\beta)/\protect\sqrt{5}$.}
\label{tab:OBE}
%\begin{tabular*}{\textwidth}{@{}l@{\extracolsep{\fill}}|rrrr}
\begin{tabular}{l|rrrr}
\hline
%%%%%%%%%%%%%%%%%%%%%%%%%%%
 {}
& $\rho$  
& $\omega_8$
&$ K^{*}$  
&Total \\ 
\hline
%%%%%%%%%%%%%%%%%%%%%%%%%%%%%%%%%%%%%%%%%%%%%%%    
{} & {} & {} & {}  \\
$ \langle  {\bf 8_a} \ {}^{1}P_1 \ | V_{ALS}| \ {\bf 8_s} \ {}^{3}P_1 \rangle$ &
8 & 4  & 8  & 20 \\
$ \langle  {\bf 8_a} \ {}^{1}P_1 \ | V_{ALS}| \ {\bf 27} \ {}^{3}P_1 \rangle$ &
6 & $-2$  & $-4$ &  0 \\
$ \langle  {\bf 10^{*}} \ {}^{1}P_1 \ | V_{ALS}| \ {\bf 8_s} \ {}^{3}P_1 \rangle$ &
$-2$ & $-2$  & 4  &  0 \\
$ \langle  {\bf 10^{*}} \ {}^{1}P_1 \ | V_{ALS}| \ {\bf 27} \ {}^{3}P_1 \rangle$ &
$-4$ & $-4$  & 8 &  0 \\
%%%%%%%%%%%%%%%%%%%%%%%%%%%%%%%%
{} & {} & {} & {}  \\
$\langle   \Lambda N^{(1/2)} \ {}^{1}P_1 \ | V_{ALS}| \  \Lambda
N^{(1/2)} \   {}^{3}P_1 \rangle$ &
0  & $-1$   & 0  & $-1$   \\
$\langle   \Sigma N^{(1/2)} \ {}^{1}P_1 \ | V_{ALS}| \ \Sigma N^{(1/2)}
\ {}^{3}P_1 \rangle$ &
2 & 1  & 0 & 3  \\
$\langle   \Sigma N^{(1/2)} \ {}^{1}P_1 \ | V_{ALS}| \ \Lambda N^{(1/2)}
\ {}^{3}P_1 \rangle$ &
$-1$  & 0  & 2  & 1  \\
$\langle   \Lambda N^{(1/2)} \ {}^{1}P_1 \ |V_{ALS}| \ \Sigma N^{(1/2)}
\ {}^{3}P_1 \rangle$ &
$-1$  & 0  & $-2$  & $-3$  \\
%%%%%%%%%%%%%%%%%%%%%%%%%%%%%%%%%%%%%%%
{} & {} & {} & {}  \\ 
$\langle  {\bf 10} \ {}^{1}P_1 \ | V_{ALS}| \ {\bf 27} \ {}^{3}P_1  \rangle$ &
1 & $-1$ & 0 & 0 \\
{} & {} & {} & {}  \\
$\langle  \Sigma N^{(3/2)} \ {}^{1}P_1 \ | V_{ALS}| \ \Sigma N^{(3/2)} \ 
{}^{3}P_1  \rangle$ &
1 & $-1$ & 0 & 0 \\
{} & {} & {} & {}  \\
\hline
\end{tabular}
\end{table}

One sees that except for the $\rep{8_s} - \rep{8_a}$ matrix element,
the sum of the $\rho$, $\omega_8$ and $K^*$ contributions vanishes
as expected from the \SU3\ symmetry.
It is also interesting to note that the difference between the $\Lam N 
\to \Lam N$ and $\Sig N \to \Sig N$ ALS matrix elements come from the
isovector $\rho$ exchange and also that the difference between 
$\Lam N \to \Sig N$ and $\Sig N \to \Lam N$ is caused by the sign
change of $K^*$ exchange.  For $\langle \Lam N \,
 ^1P_1| V_{ALS}|\Sig N^{(1/2)} \,^3P_1\rangle$, 
 the $\rho$ and $K^*$ exchanges 
are added up while they tend to cancel for
$\langle\Sig N^{(1/2)} \, ^1P_1| V_{ALS}|\Lam N\,^3P_1\rangle$.

\subsection{\SU3\ Breaking}
When the \SU3\ symmetry breaking is taken into account, the
coefficients in Table \ref{tab:OBE} for 
$\rho$, $\omega_8$ and $K^*$ cannot simply be summed up, as the
interaction range for $K^*$ is significantly shorter than that for 
$\rho$ or $\omega_8$.
Furthermore the differences among the octet baryon masses generate
various terms that would vanish in the \SU3\ limit.  The full form of
the ALS potential for the scalar, pseudoscalar and vector exchanges
are given in the following.
\begin{equation} 
V^{ALS}_{PS}(r) =   \label{eqn:ALS-PSEUDO}
  g_{13}g_{24} \frac{m^3}{4 \pi} \xi(x)\frac{1}{8}
\left(\frac{1}{M_1 M_4} -\frac{1}{M_2 M_3}\right) 
   {\vec L} \cdot ( \vsig_1 - \vsig_2 ) P_{\sigma} 
\end{equation}
\begin{equation} 
V^{ALS}_{S}(r) =     \label{eqn:ALS-SCALAR}
-  g_{13}g_{24} \frac{m^3}{4 \pi} \xi(x)\frac{1}{8}
\left(\frac{1}{M_1 M_3} -\frac{1}{M_2 M_4}\right) 
   {\vec L} \cdot ( \vsig_1 - \vsig_2 )  
\end{equation}
\begin{eqnarray}
V^{ALS}_{V}(r) & = & 
-\frac{m^3}{4 \pi} \xi(x) {1\over 2}\,
{\vec L} \cdot ( \vsig_1 - \vsig_2 )
\nonumber \\
&\times & 
\left[ \left(g_{13}+ \frac{M_1+M_3}{2 \cal M}f_{13}\right)
       \left(g_{24}+ \frac{M_2+M_4}{2 \cal M}f_{24}\right) 
\frac{1}{4} \left( \frac{1}{M_2 M_3} -  \frac{1}{M_1 M_4}\right)P_{\sigma}
               \right.  \nonumber \\
&- & 
 \frac{g_{13}f_{24} -f_{13}g_{24}}{4 \cal M}
  \left( \frac{1}{M_1}+ \frac{1}{M_2}+ \frac{1}{M_3}+ \frac{1}{M_4} \right)
         \nonumber \\
&+ &
  g_{13}g_{24} \frac{1}{4}
 \left( \frac{1}{M_1 M_3} - \frac{1}{M_2 M_4} \right) 
 \left( 1+ {\frac{(M_3-M_1)(M_4-M_2)}{m^2}} \right)\nonumber \\
&+ &
  \frac{f_{13}f_{24}}{4 {\cal M}^2}
\left\{ \frac{1}{2} \left( \frac{M^2_2 + M^2_4}{M_2 M_4}
                   -\frac{M^2_1 + M^2_3}{M_1 M_3} \right) \right.\nonumber \\
&- &
m^2  \left\{
\frac{1}{16} \left( \frac{1}{M_1 M_3}-\frac{1}{M_2 M_4} \right)
  \right.          \nonumber \\
&+ &  \frac{1}{32}  \left(
\frac{1}{M_1^2} + \frac{1}{M_3^2}
-\frac{1}{M_2^2} - \frac{1}{M_4^2}
+  \frac{M_1}{M^3_3} + \frac{M_3}{M^3_1}
 -\frac{M_2}{M^3_4} - \frac{M_4}{M^3_2} \right. \nonumber \\
&+ &\left. \left. \left.
 \frac{ M^2_2 + M^2_4 - M^2_1 - M^2_3 }{M_1 M_2 M_3 M_4}
\right)  \right\} \right] 
\end{eqnarray}
where the scattering of $M_1 + M_2 \to M_3 + M_4$ is considered,
while $\cal M$ is the mass of the proton.
The terms that contains
the spin exchange operator,
\begin{equation}
  P_{\sigma} = {1+ \vsig_1\cdot\vsig_2\over 2}
\end{equation}
come from the exchange part in the $K$ and $K^*$ exchanges.

Fig.\ 1 shows the ALS part of the one-boson exchange potential
(OBE) and the adiabatic potential in the quark cluster model (QCM) for 
the $I=3/2$ $\Sigma N$ channel, \ie,
$\langle  \Sigma N^{(3/2)} \ {}^{1}P_1 \ | V_{ALS}| \ \Sigma N^{(3/2)} \ 
{}^{3}P_1  \rangle$.
It can be compared with the corresponding ordinary LS (SLS) part for 
$\langle  \Sigma N^{(3/2)} \ {}^{3}P_1 \ | V_{SLS}| \ \Sigma N^{(3/2)} \ 
{}^{3}P_1  \rangle$, shown in Fig.\ 2.
The curves in these figures show various cases in \SU3\ breaking.
The potentials in the \SU3\ symmetric limit are
labeled by ``Bs-Ms'' for OBE and ``QCM-s'' for QCM.
The ALS potentials are zero in this limit.
The symmetry is broken in ``QCM-d'' by the quark mass difference,
while in ``Bd-Md'' both the meson mass differences and the baryon mass
differences are taken into account.
It is clear that the ALS is strongly suppressed
even when the \SU3\ symmetry breaking is considered.
\SU3\ breaking effect is small also in SLS.
In general, we observe that the ALS potential is very weak for 
the two-baryon channels in which the ALS vanishes in the \SU3\ limit.
In these figures, potential parameters are chosen from the Nijmegen 
model D potential in OBE and ref.{\cite{Morimatsu}} for QCM.

Figs.\ 3 and 4 shows the OBE and QCM adiabatic potentials in the
$\Lam N (^1P_1)\to \Sig N^{(1/2)} (^3P_1)$ (Fig.~3), and 
$\Sig N^{(1/2)} (^1P_1)\to \Lam N(^3P_1)$ (Fig.~4). 
They show that the ALS potentials are rather strong.
It is also indicated that the QCM gives stronger ALS than OBE and
also that the effects of \SU3\ breaking is less for QCM.

\begin{figure}[htbp]
\begin{minipage}[t]{77mm}
\begin{center}
\epsfxsize=75mm
\epsfbox{file1.eps}
\caption{The OBE and QCM potentials in 
$\langle  \Sigma N^{(3/2)} \,^{1}P_1 \, | V_{ALS}| \, 
\Sigma N^{(3/2)} \,^{3}P_1  \rangle$.}
\end{center}
\end{minipage}
\hspace{\fill}
\begin{minipage}[t]{77mm}
\begin{center}
\epsfxsize=75mm
\epsfbox{file2.eps}
\caption{The OBE and QCM potentials in 
$\langle  \Sigma N^{(3/2)} \,^{3}P_1 \,| V_{SLS}| \,
\Sigma N^{(3/2)} \,^{3}P_1  \rangle$.}
\end{center}
\end{minipage}
\end{figure}

\begin{figure}[htbp]
\begin{minipage}[t]{77mm}
\begin{center}
\epsfxsize=75mm
\epsfbox{file3.eps}
\caption{The OBE and QCM potentials in 
$\langle \Lambda N^{(1/2)} \,^{1}P_1 \, |V_{ALS}| \, \Sigma N^{(1/2)}
\,^{3}P_1 \rangle$. }
\end{center}
\end{minipage}
\hspace{\fill}
\begin{minipage}[t]{77mm}
\begin{center}
\epsfxsize=75mm
\epsfbox{file4.eps}
\caption{The OBE and QCM potentials in 
$\langle \Sigma N^{(1/2)} \,^{1}P_1 \,| V_{ALS}| \, \Lambda N^{(1/2)}
\,^{3}P_1 \rangle$. }
\end{center}
\end{minipage}
\end{figure}

%
%\begin{figure}[htbp]
%\begin{minipage}[t]{77mm}
%\begin{center}
%\epsfxsize=75mm
%\epsfbox{file5.eps}
%\caption{Recoupling of SU(3) state vectors.}
%\end{center}
%\end{minipage}
%
%\hspace{\fill}
%
%\begin{minipage}[t]{77mm}
%\begin{center}
%\epsfxsize=75mm
%\epsfbox{file6.eps}
%\caption{Recoupling of SU(3) state vectors.}
%\end{center}
%\end{minipage}
%\end{figure}

\section{Conclusion}

We have studied the \SU3\ symmetry of the ALS interactions in the YN force,
and found that the \SU3\ symmetry is rather good in accounting the 
properties of ALS forces in various baryon channels.
The YN ALS forces due to the quark exchange are significantly large, 
comparable to the ordinary LS force of the same origin.
The meson exchange force almost vanishes except for a term proportional 
to the difference in $F/D$ ratios of the vector and tensor couplings 
of the vector mesons.  Thus the ALS interaction has a shorter range
than the ordinary LS force, or the tensor forces.
It is extremely interesting to pin down the strengths and the properties
of the ALS forces in the YN sector in determining the origin of the 
baryonic forces.  
Further theoretical and experimental studies of the YN spin-orbit 
interactions are very much encouraged.

\bigbreak
The author acknowledges Y.~Tani, K.~Yazaki, O.~Morimatsu and 
S.~Takeuchi for useful discussions.  
This work is supported in part by the Grant-in-Aid for scientific
research (C)(2)08640356 and Priority Areas (Strangeness Nuclear
Physics) of the Ministry of Education, Science and Culture of Japan.

\def \vol(#1,#2,#3){{{\bf {#1}} (19{#2}) {#3}}}
\def \NP(#1,#2,#3){Nucl.\ Phys.\          \vol(#1,#2,#3)}
\def \PL(#1,#2,#3){Phys.\ Lett.\          \vol(#1,#2,#3)}
\def \PRL(#1,#2,#3){Phys.\ Rev.\ Lett.\   \vol(#1,#2,#3)}
\def \PRp(#1,#2,#3){Phys.\ Rep.\          \vol(#1,#2,#3)}
\def \PR(#1,#2,#3){Phys.\ Rev.\           \vol(#1,#2,#3)}
\def \PTP(#1,#2,#3){Prog.\ Theor.\ Phys.\ \vol(#1,#2,#3)}
\def\MO{M.~Oka} \def\KY{K.~Yazaki}
\def \ibid(#1,#2,#3){{\it ibid.}\         \vol(#1,#2,#3)}
\def\MOka{\MO}
\def\KYazaki{\KY}
\def\Sachiko{\ST}
\def\PRD(#1,#2,#3){\PR(D#1,#2,#3)}
\def\PRC(#1,#2,#3){\PR(C#1,#2,#3)}
\def\etal{{\it et al.}}


\begin{thebibliography}{9}

\bibitem{YITP}
\MO, Prog.\ Theor.\ Phys.\ Supple.\ \vol(120,95,95);
J.\ Korean Phys.\ Soc., \vol(29,96,S326).

\bibitem{TO} 
Y.~Tani and \MOka, to be published.

\bibitem{ITO} 
T.~Inoue, S.~Takeuchi and \MO, \NP (A597,96,563);
T.~Inoue, \MO, T.~Motoba and K.~Itonaga, preprint:
		``Non-mesonic Weak Decays of Light Hypernuclei in the Direct Quark
			and the One Pion Exchange Mechanisms''.

\bibitem{SOS} 
K.~Saito, \MO\ and T.~Suzuki, preprint:
		``Exchange Currents for Hypernuclear Magnetic Moments'',
		nucl-th/9705029,  April, 1997.


\bibitem{Nijmegen}
M.M.~Nagels, T.A.~Rijken and J.J.~de Swart, \PR(D12,75,744);
\ibid(15,77,2547);  \ibid(20,79,1633); \ibid(17,78,768);
P.M.M.~Maessen, Th.A.~Rijken and J.J.~de Swart, \PR(C40,89,2226).

\bibitem{DF}
C.B.~Dover and H.~Feshbach,  Ann.\ Phys.(NY)\vol(198,90,321);
\ibid(217,92,51).
 
\bibitem{Julich}
 B.~Hilzenkamp, K.~Holinde and J.~Speth, \NP(A500,89,485);
 A.~Reuber, K.~Holinde and J.~Speth, \NP(A570,94,543).

\bibitem{Bonn}
R.~Machleidt, K.~Holinde and Ch.~Elster, \PRp(149,87,1).

\bibitem{OSY} 
% \bibitem{QCMYN}
\MOka, K.\ Shimizu and \KYazaki, \PL(B130,83,365);
\MOka, K.\ Shimizu and \KYazaki, \NP(A464,87,700).

\bibitem{NSF}
Y.~Fujiwara, C.~Nakamoto and Y.~Suzuki, \PTP(94,95,215);
\ibid(94,95,353);	\PRL(76,96,2242).

\bibitem{Jaffe}
R.L.~Jaffe, \PRL(38,77,195).

\bibitem{hakone} 
\MO, Hyperfine Interactions, \vol (103,96,275).

\bibitem{AP-exp}
R.~Abegg, \etal, \PRL(56,86,2571); \PRD(39,89,2464);
L.D.~Knutson, \etal, \PRL(66,91,1410); S.E.~Vigdor, \etal, 
\PRC(46,92,410).

\bibitem{OY}
M.\ Oka and K.\ Yazaki, \PL(B90,80,41);
             \PTP(66,81,556); \ibid(66,81,572);
         in {\sl Quarks and Nuclei}, ed.\ by W.\ Weise
                                (World Scientific, 1985);
K.\ Shimizu, Rep.\ Prog.\ Phys.\ \vol(52,89,1).
  

\bibitem{Morimatsu}
O.~Morimatsu, S.~Ohta, K.~Shimizu, and \KY, \NP(A420,84,573).

%% 
 % \bibitem{NST}
 % V.G.\ Neudatchin, Yu.F.\	Smirnov	and	R.\	Tamagaki, \PTP(58,77,1072).
%\bibitem{OSO}
%K.~Ogawa, S.~Takeuchi and \MO, 
%		in {\sl Properties and Interactions of Hyperons},
%		ed.\ by B.F. Gibson, P.D.\ Barnes and K. Nakai,
%		p.169,  (World Scientific, 1994);
%	to be published.
%
%\bibitem{Tubingen}
%F.~Fern\'andez, A.~Valcarce, U.~Straub and A.~Faessler,
%Jour.\ Phys.\ G\vol(19,93,2013);
%Z.-Y.~Zhang, A.~Faessler, U~Straub and L.Ya.~Glozman,
%NP(A578,94,573).

\end{thebibliography}
\end{document}